\documentclass[times, twoside, watermark]{zHenriquesLab-StyleBioRxiv}

\leadauthor{Materne} 

\title{Effective strains enable rapid wound closure in jellyfish after injury}

\shorttitle{Jellyfish wound closure}

\author[1,2]{Anne Materne}
\author[1,2,3]{Zhiqi Shen}
\author[4]{Chiara Sinigaglia}
\author[1,2,5,*]{Carl D. Modes}

\affil[1]{Max Planck Institute of Molecular Cell Biology and Genetics, Pfotenhauerstr. 108, 01307 Dresden, Germany}
\affil[2]{Center for Systems Biology Dresden, Pfotenhauerstr. 108b, 01307 Dresden, Germany}
\affil[3]{Southern University of Science and Technology, Shenzhen, China}
\affil[4]{CNRS, Sorbonne Université, Biologie Intégrative des Organismes Marins (BIOM), Banyuls-sur-Mer, France}
\affil[5]{Physics of Life Cluster of Excellence, TU Dresden, Arndoldstr. 18, 01307 Dresden, Germany}
\affil[*]{modes@mpi-cbg.de}

\usepackage{graphicx} 
\usepackage{amsmath, amsfonts}
\usepackage{color}
\usepackage{gensymb}
\usepackage{caption}
\usepackage{cuted}
\usepackage{xurl}

\begin{document}

\maketitle

\begin{abstract}
    The jellyfish \textit{Clytia hemisphaerica} possesses astounding regenerative capacities and is able to close even large wounds within a few hours. This rapid pace of wound closure raises the question whether tissue mechanics, rather than tissue restructuring or cell proliferation, might be underlying the process. We tested this possibility by asking if simple pre-strains within the jellyfish umbrella would be capable of initiating wound closure in a jellyfish body geometry. To this end, we employed an \textit{in silico} spring lattice model, a coarse-grained model of elastic materials which has previously been established to study tissue mechanics problems. We found that, using radially contractile (but not radially extensile) strains, wound closure can indeed be initiated across a wide range of conditions. This is even true for large cut sizes and, hence, small pieces of remaining tissue material, in good agreement with the experimental findings. Finally, we derived an analytical expression for the expected amount of achievable closure as a function of the residual material angle. These results establish important foundations for further investigations of the biophysics underpinning jellyfish regeneration. 
\end{abstract}

\section*{Introduction} \label{sec:Introduction}

\begin{figure*}
    \centering
    \includegraphics[width=\textwidth]{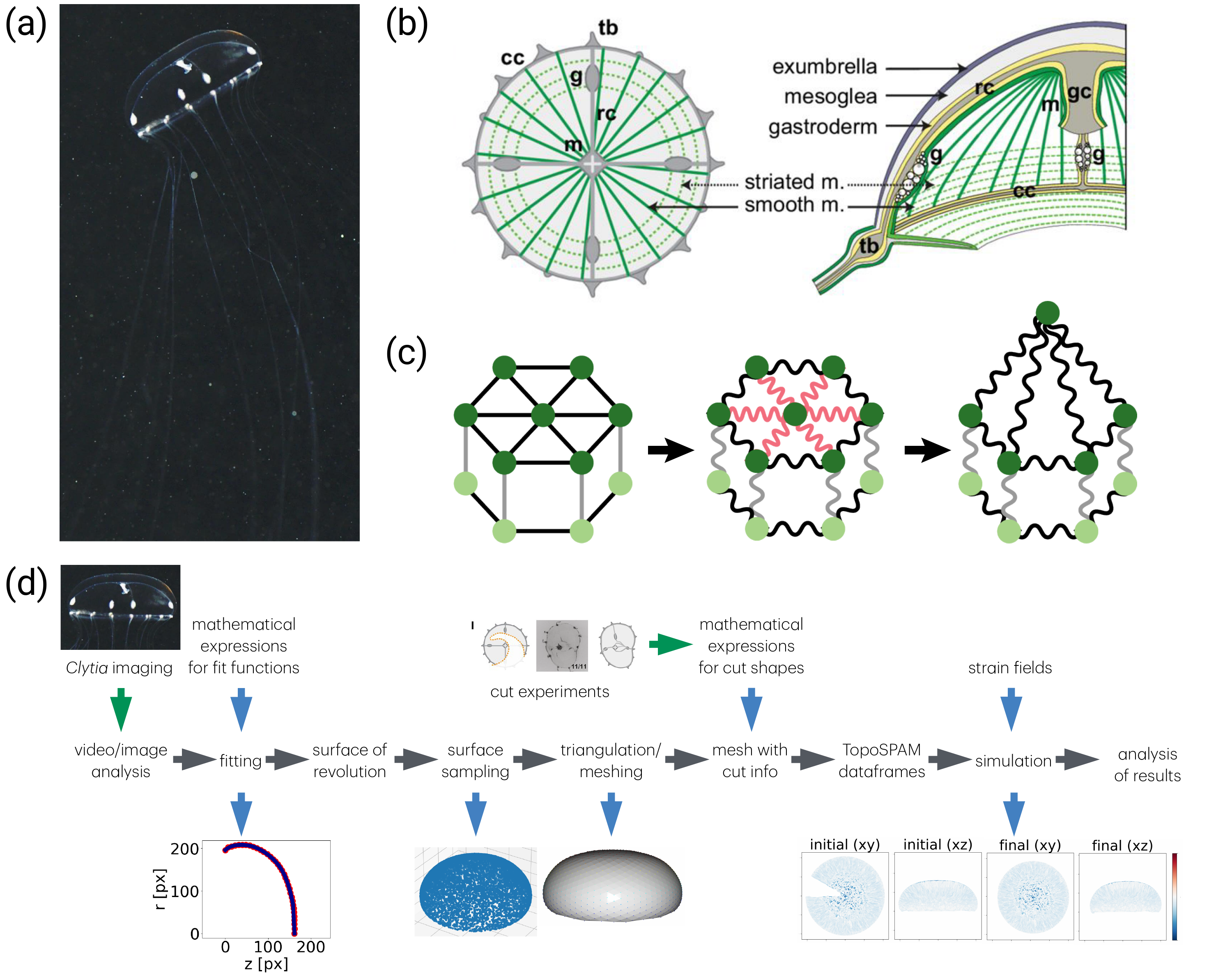}
    \caption{(a) Video capture image of a swimming (uninjured) \textit{Clytia hemisphaerica} jellyfish. 
    (b) Anatomical sketches of the \textit{Clytia hemisphaerica} umbrella (left: top view, right: cross section) with solid and dashed green lines indicating smooth and striated muscle arrangement, respectively, and with cc = circular canal, g = gonad, gc = gastric cavity, m = manubrium, rc = radial canal, tb = tentacle bulb. Image adapted from~\cite{Sinigaglia2020}. 
    (c) Schematic of a spring lattice model. Surfaces are represented by points in 3D (here different colours indicate points on different surfaces), which can be connected in (black lines) or across layers (grey lines). In the model, connections are springs, which endow the system with mechanical properties. In the initial state (middle panel), individual springs can be in their relaxed state or under strain (i.e. compressed or extended, shown in red). The simulation aims to relax all the springs and to this end moves the points, thus deforming the surfaces (right-most panel). 
    (d) Schematic of the whole modelling procedure, including generation of a mathematical description of the mesoglea shape, surface sampling, meshing, introduction of cuts into the mesh, set-up of the simulation procedure, introduction of strain fields, simulation runs, and data analysis. The part of this figure labelled `cut experiments' has been adapted from~\cite{Sinigaglia2020}.}
    \label{fig:intro_problem_setup}
\end{figure*}

Jellyfish are multicellular animals with a comparatively simple anatomy (Figures~\ref{fig:intro_problem_setup}(a) and (b)). A jellyfish's umbrella -- the main part of its medusoid body -- only consists of a few cellular tissue layers enclosing a thick layer of extracellular matrix (ECM) called the mesoglea (Figure~\ref{fig:intro_problem_setup}(b)). The mesoglea plays a key role in the medusa. It acts as a hydrostatic skeleton and counteracts the muscle contractions during the recovery phase of a swimming stroke~\cite{Alexander1964, Weber1985, Gambini2012}. 

Furthermore, within the Cnidarians, the group of animals to which the medusae belong, species with astounding regenerative capacities can be found. 
For example, the hydrozoan jellyfish \textit{Clytia hemisphaerica} is able to close large wounds rapidly (within a few hours) and can even regrow lost organs~\cite{Sinigaglia2020, Schmid1976, Kamran2017}. 
\textit{Clytia} shares this regenerative capacity with its close relative \textit{Hydra vulgaris}, which is also being studied as a regeneration model by some labs~\cite{Bergheim2019, Maroudas-Sacks2025, Bailles2025, Cox2025}. Although both species belong to the same class of animals, major differences exist between \textit{Hydra} and \textit{Clytia}, and in their respective regeneration. 
For example, the medusoid form of \textit{Clytia} exhibits a very different body geometry and anatomy than the \textit{Hydra} polyp. Additionally, the thick jellyfish mesoglea layer does not exist in \textit{Hydra}, whose body only consists of two epithelial cell layers with a comparatively thin ECM in between~\cite{Bergheim2019}. 
Finally, a medusa is a free swimming animal while a polyp is sessile. Thus, these animals are subject to distinct physical constraints, such as those imposed by the surrounding fluid flows. \\

Given the pace of wound closure in \textit{Clytia}, from a biophysical perspective the question arises: how might spontaneous strains emerging within the animal's umbrella after injury contribute to this process? A mechanically induced wound closure has the potential to be much faster than tissue restructuring or cell proliferation. Based on the requirements for shape maintenance and swimming, it is very likely that strains exist in the real, biological system. However, the details of such physiological strains remain unknown. 

We here took inspiration from the possibility of mechanical strains contributing to wound closure and began by asking whether such strains alone are even capable of closure initiation in the \textit{Clytia} body geometry.
To this end, we built a simple \textit{in silico} representation of the \textit{Clytia} umbrella. We employed a spring lattice model, a coarse-grained model of elastic materials (depicted in Figure~\ref{fig:intro_problem_setup}(c)), which has successfully been applied to tissue mechanics problems in the past~\cite{Singh2025, Ramos2025, Santoriello2026}.
In such a model, a surface is represented by a number of points (dark-green and light-green dots in Figure~\ref{fig:intro_problem_setup}(c) for two different surfaces). Thus, points do not necessarily symbolise actual surface features such as cells, but rather sample the overall surface geometry. The model points are connected by springs (black lines in Figure~\ref{fig:intro_problem_setup}(c) left panel), which endow the surface with elastic properties. Cross-connections between different layers are similarly possible (grey lines in Figure~\ref{fig:intro_problem_setup}(c) left panel). The springs themselves can be in a relaxed state or be under strain (meaning compressed or extended, marked in red in Figure~\ref{fig:intro_problem_setup}(c) middle panel) at the start of the simulation. The simulation aims to relax the entire set of springs as much as possible by moving the points in 3D space, hence deforming the surface (Figure~\ref{fig:intro_problem_setup}(c) right panel). 
 
We here focus on a set of simple, yet rich example strains in the radial and circumferential directions, in line with the underlying geometry and anatomy of the jellyfish body.
We analysed their potential for wound closure initiation depending on the following factors: whether the strains are contractile or extensile, the magnitudes of the strain at the jellyfish umbrella apex and widest parts (close to the rim), the strain field variation in between these two positions, the type of cut (here: straight vs. spiral cut lines) and the cut opening angle. 

We found that wound closure can indeed be initiated purely mechanically in a wide variety of conditions. Further, we quantified the amount of closure as a function of the underlying model parameters and cut shapes. Finally, we present an analytical expression for the amount of closure as a function of residual tissue material. 

\section*{Results} \label{sec:Results}

\subsection*{Radially contractile but not radially extensile strain fields initiate wound closure in a simple model of the jellyfish umbrella}

\begin{figure*}
    \centering
    \includegraphics[width=0.94\textwidth]{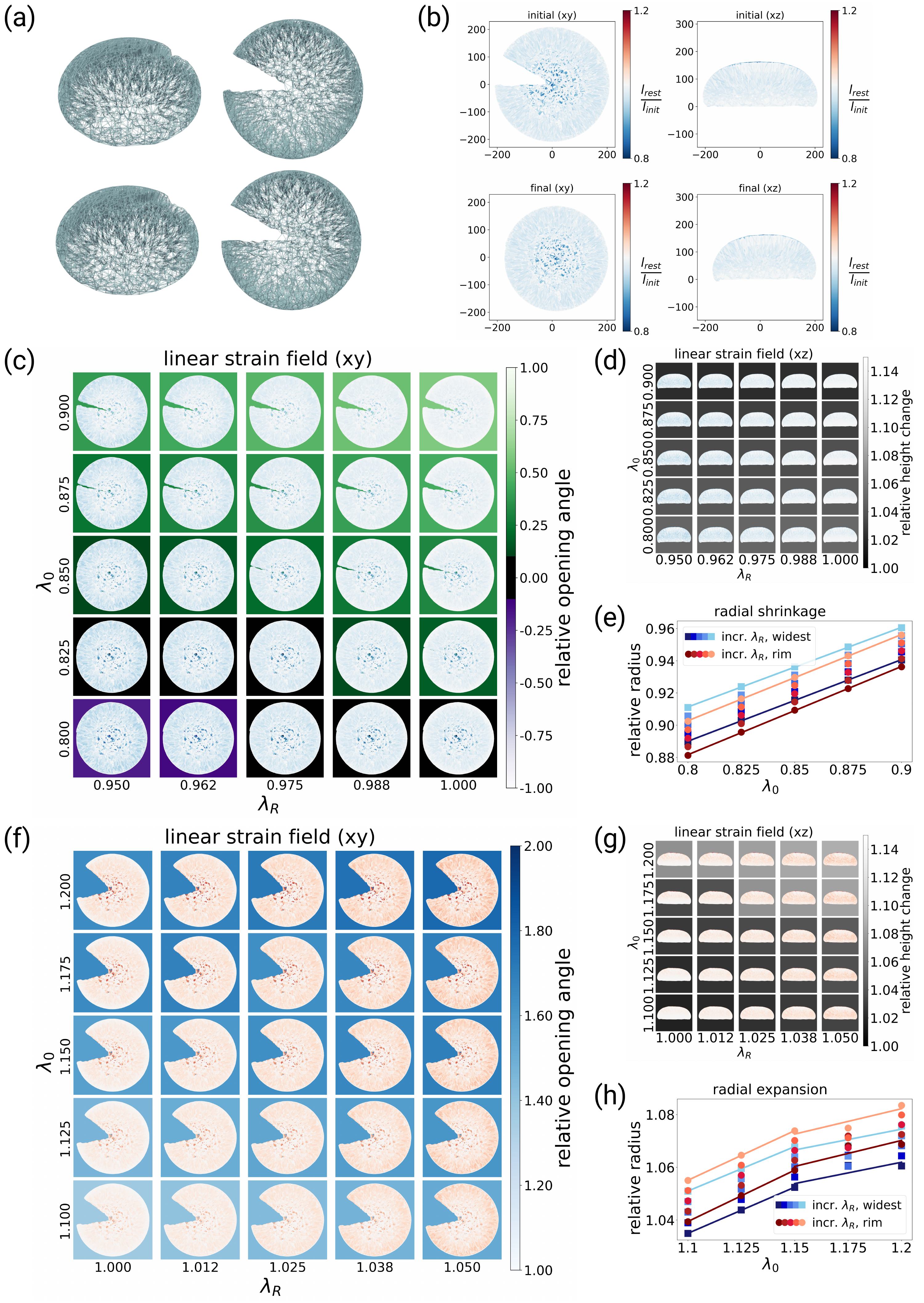}
    \caption{caption see next page}
\end{figure*}

\begin{figure*}[ht]
    \ContinuedFloat 
    \centering
    \caption{
    (a) Two different simulation meshes (initial states), both with a 30\degree linear cut (side and top view each). The mesh shown in the top line has been used for the simulations presented in this figure. The other mesh has been used for repeat-runs, e.g. those shown in Figure~\ref{fig:sweeps_and_metrics}.
    (b) A typical simulation result. The shown run was performed with a linearly varying ($p=1$), radially contractile strain field, $\lambda_{0} = 0.825$, $\lambda_{R} = 0.975$. The initial (top row) and final states (bottom row) of the simulation mesh are shown in side- and top view each. The color bar indicates the magnitude of the applied strain field (initial state) and residual strains (final state), respectively.
    (c) Simulation results using a set of linearly varying, radially contractile strain fields. The x- and y-axes indicate the values of $\lambda_{R}$ and $\lambda_{0}$, respectively, for each simulation. Each subplot shows the top view of the simulation's final state in the foreground. The background colour of each subplot encodes the relative change of the opening angle between the final and initial states, as indicated by the colour bar.
    (d) Results of the same simulations as in (c). Here, the side view of the simulations' final states are shown in the foreground and the relative height change between the final and initial states (as indicated by the colour bar) is depicted by the background colour. 
    (e) Results of the same simulations as in (c). Here, the relative radius change between final and initial states is shown as a function of $\lambda_{0}$. Blue squares indicate the change at the widest part of the mesoglea, red circles the change at the rim. Lines represent linear fits through individual sets of data points with the same $\lambda_{R}$-value. 
    (f) Simulation results using a set of linearly varying, radially extensile strain fields, as the values of $\lambda_{R}$ and $\lambda_{0}$ on the axes indicate (simulation mesh and other parameters as in (c)). Note that, compared to (c), the background colour bar has changed.  
    (g) Results of the same simulations as in (f), but depicted are the side views of the simulation meshes' final states and the relative height changes, as in (d). 
    (h) Results of the same simulations as in (f), but depicted are the relative radius changes at the widest part of the mesoglea and the rim as in (e). In contrast to (e), the relative radii appear not to vary strictly linearly with $\lambda_{0}$. Therefore, for each set with the same $\lambda_{R}$-value, two linear fits have been performed, including the first and last three data points, respectively.}
    \label{fig:linear_cut}
\end{figure*}

First, we aimed to establish whether it is possible at all that simple, radially and circumferentially directed strains initiate wound closure in jellyfish. 
To this end, we built an idealised, 3-dimensional \textit{in silico} model of the \textit{Clytia} umbrella (see Figure~\ref{fig:intro_problem_setup}(c), as well as Methods ``Geometric description of an idealised jellyfish body'' and ``Point sampling and mesh generation''). Briefly, we outlined the cross-sectional shapes of the exumbrella and subumbrella tissues in a video capture image of an uninjured \textit{Clytia} medusa (compare also Figure~\ref{fig:intro_problem_setup}(a)), fitted polynomial functions to these outlines, generated surfaces of revolution from the fits, sampled points randomly and uniformly on these surfaces and triangulated those points to obtain a simulation mesh. Different meshes for repeat-simulations were generated from the same polynomial functions and surfaces of revolution thereof, but using new, independent sets of points sampled on these surfaces.  
This \textit{in silico} representation of the jellyfish umbrella is idealised because we did not distinguish any individual tissue layers and considered the (uncut) umbrella to be rotationally symmetric, omitting the existence of canals and gonads, for example.

In line with previous experimental work~\cite{Sinigaglia2020}, straight or spiral cuts were introduced to the simulation meshes after triangulation (see Methods ``Point sampling and mesh generation''). 
Figure~\ref{fig:linear_cut}(a) shows two of the resulting, cut simulation meshes in isometric projection and top-views each. For the mesh in the top line, we have run a wide range of simulations, to determine under which conditions wound closure might be initiated. The second mesh (together with others not shown here) has been used to run repeat simulations, for example to generate the averages shown in Figure~\ref{fig:sweeps_and_metrics}.

As the physiological strains in the jellyfish umbrella remain unknown, we focused on a set of simple, yet rich example strain fields. Following the procedure described in~\cite{Fuhrmann2024}
\begin{equation}
\boldsymbol{\lambda} (\boldsymbol{x})=\lambda_{rr}(\mathbf{e}_r  \otimes \mathbf{e}_r)+\lambda_{\theta \theta}(\mathbf{e}_{\theta} \otimes \mathbf{e}_{\theta})+\lambda_{zz}(\mathbf{e}_z  \otimes \mathbf{e}_z) ,
\label{eqn:strain_tensor_field}
\end{equation}
is the pre-strain tensor field with $\boldsymbol{x}$ denoting the reference position of a mesh point in 3D cylindrical coordinates $(\mathbf{e}_r, \mathbf{e}_{\theta}, \mathbf{e}_{z})$. The $z$-axis is defined by the symmetry axis through the umbrella apex and we positioned the origin at the centre of the umbrella rim. Thus, the diagonal entries $\lambda_{ii}(\boldsymbol{x})$ are the strain field components along the $\mathbf{e}_i$ direction.
Based on the rotational symmetry of the idealised jellyfish representation, we focussed on strains that depend on the radial coordinate $r$ only (note that in cylindrical coordinates $r$ is the projected radial distance in the $xy$-plane) and here studied fields of the form 
\begin{equation}
\lambda_{rr}(r)=(\lambda_{R} -\lambda_{0})\left( \frac{r}{R} \right)^{p} + \lambda_{0} \, ,
\label{eqn:radial_strain_field}
\end{equation}
and $\lambda_{\theta \theta}=\lambda_{zz}=1$. 
In this expression, $R=r_{\mathrm{max}}$ is the radius of the widest part of the umbrella and $\lambda_{R}$ and $\lambda_{0}$ are the magnitudes of the strain at the widest part and at the apex, respectively. Finally, the power $p$ defines how the strain field varies in between the values $\lambda_{0}$ and $\lambda_{R}$. Details on how these strain fields are being applied to the mesh can be found in the Methods section ``Strain fields''.
If $\lambda_{rr} < 1$, the strain field is radially contractile, otherwise, if $\lambda_{rr} > 1$, it is radially extensile. 
It should be noted that radially contractile strain fields generally lead to circumferential extension of the material. Conversely, radially extensile strains generally lead to circumferential contraction and circumferential strains also lead to radial changes. This is due to the Poisson effect, describing the perpendicular deformation of a material under load. \\

We began our investigation with the simplest possible wound shape, namely two linear cuts from the umbrella apex to the rim with an opening angle $\alpha$ (here we chose $\alpha_0 = 30\degree$, that is $\alpha \gtrapprox 30\degree$) and a linearly varying strain field, $p = 1$. 
The outcome of a single simulation run with a given strain field is exemplified in Figure~\ref{fig:linear_cut}(b) (see also Methods ``Simulation runs''). In the top row, the initial state of the simulation is presented in both top (left image) and side view (right image), while the bottom row shows the same views but for the final state of the simulation. The colours on the mesh indicate the magnitude of the strain at each location according to the provided colour bar. This colour bar is consistent for all figures that depict strain fields throughout this work. Note that there might be residual strains in the system even when a simulation has finished running, as is the case here (indicated by the mesh colouring in the final state). 

Figure~\ref{fig:linear_cut}(c) shows the final states of a entire set of simulation runs for radially contractile strain fields with different strain values $\lambda_{0}$ and $\lambda_{R}$ (and with $p=1$). The two $\lambda$-values of each individual run are provided on the figure axes. 
Each subplot shows the top view of the simulation's final state in the foreground. The background colour indicates the amount of wound closure determined by the relative opening angle of the cut between the final and initial stages, $\alpha_{\text{final}} / \alpha_{\text{initial}}$ (see also Methods ``Quantification of wound closure'').
As can be seen from this image, the relative opening angle is smaller than $1$ for all the performed simulations.
This indicates that wound closure is being initiated in all of these cases. 

The results of the same simulations are also presented in Figure~\ref{fig:linear_cut}(d), but here, plots of the front views of each simulation's final state are depicted in the foreground. Additionally, the background colour indicates the relative height of the simulation structure between final and initial states, $h_{\text{final}} / h_{\text{initial}}$ (see also Methods ``Quantification of mesh shape change''), indicating slight increases in height for all the simulations.  

Furthermore, Figure~\ref{fig:linear_cut}(e) depicts the relative radial changes of the simulated meshes as a function of $\lambda_{0}$, the strain value at the umbrella apex. Blue squares indicate the relative radii at the widest part of the umbrella, while red circles indicate the change at the umbrella rim. In each case, darker to lighter colours represent simulations with increasing values of $\lambda_{R}$. 
The lines are linear fits to the data for the smallest and largest $\lambda_{R}$-values each.
As this plot shows, the relative radii appear to depend linearly on both $\lambda_{0}$ and $\lambda_{R}$ (although we did not fit the latter relationship). \\

In contrast to the above, Figures~\ref{fig:linear_cut}(f)-(h) show simulation runs with radially extensile strains. 
As the change in background colour bar between Figures~\ref{fig:linear_cut}(c) and (f) indicates, the relative opening angles are larger than $1$ in the radially extensile cases.
Thus, in contrast to their radially contractile counter parts, the strain fields tested here do not initiate wound closure. Instead, they open up existing wounds even further. 

Interestingly, it appears that both contractile and extensile strain fields lead to slight increases in overall structure height (compare Figures~\ref{fig:linear_cut}(d) and (g)). 
Also, while the relative radii at the umbrella rim and widest parts appear to be linearly dependent on $\lambda_{0}$ in the case of radially contractile strain fields, this does not seem to be the true for radially extensile fields. 
Instead, the simulation results in Figure~\ref{fig:linear_cut}(h) show a linear regime only for weakly extensile fields. 

Because we are interested in the conditions that lead to wound closure initiation in this work, in the following we exclusively focus on radially contractile strains.

\subsection*{The strain field variation influences the amount of closure for given ($\lambda_{0}$,$\lambda_{R}$)-values}

\begin{figure*}
    \centering
    \includegraphics[width=0.99\textwidth]{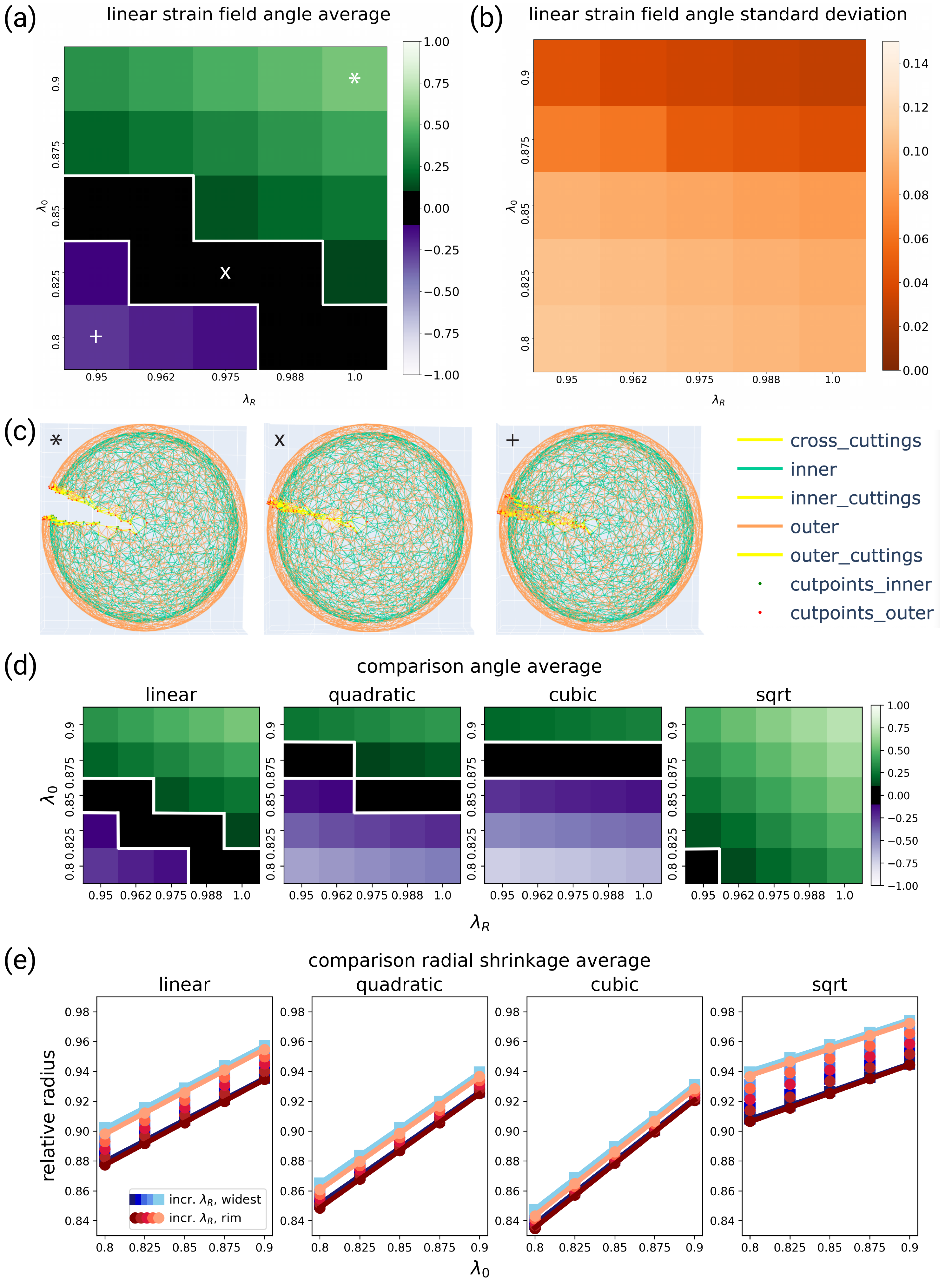}
    \caption{caption see next page} 
\end{figure*}

\begin{figure*}[ht]
    \ContinuedFloat 
    \centering
    \caption{
    (a) Relative change of the opening angle (as indicated by the colour bar) averaged over three sets of simulations runs similar to the set shown in Figure~\ref{fig:linear_cut}(c). For each set of runs, a different simulation mesh has been used, as indicated in Figure~\ref{fig:linear_cut}(a). The background colours reflect different types of simulation outcomes in terms of closure levels (green = under-closure, black = proper closure, purple = over-closure) and the white lines highlight the phase boundaries between these types.
    (b) Standard deviations of the averaged relative opening angle values shown in (a). 
    (c) Examples of the different possible closure levels from under-closed (left panel), via proper- (middle panel) through over-closed (right panel). Each panel depicts the final state of a simulation run with the set of parameters indicated by the corresponding symbols in (a) and using the mesh also used for Figure~\ref{fig:linear_cut}. 
    (d) Comparison of average amounts of closure for four parameter sweeps using radially contractile strain fields with different powers of $r$ (averaged over three runs with different simulation meshes each). The respective powers are $p=1$ for linear (same data as in (a)), $p=2$ for quadratic, $p=3$ for cubic, and $p=1/2$ for sqrt. 
    (e) Comparison of relative radius changes for the same four parameter sweeps as in (d) (similarly averaged over three runs with different simulation meshes each). As in Figure~\ref{fig:linear_cut}(e), blue squares indicate the change at the widest part of the mesoglea, red circles the change at the rim and lines represent linear fits through sets of data points with the same $\lambda_{R}$-value.}
    \label{fig:sweeps_and_metrics}
\end{figure*}

An important additional observation that we can make from Figure~\ref{fig:linear_cut}(a), is that the amount of achieved closure is not identical for all the tested pairs of ($\lambda_{0}$, $\lambda_{R}$)-values. 
We therefore now analyse the different levels of closure in more detail.

Figure~\ref{fig:sweeps_and_metrics}(a) shows similar results as Figure~\ref{fig:linear_cut}(c), but here, three runs with different simulations meshes (including both meshes shown in Figure~\ref{fig:linear_cut}(a)) have been averaged (colour bar same as in Figure~\ref{fig:linear_cut}(c), see also Methods ``Averaging over multiple simulation runs''). 

The corresponding standard deviations are shown in Figure~\ref{fig:sweeps_and_metrics}(b). That the values of the standard deviation are small indicates that the simulation outcome does not depend significantly on the chosen mesh.

In the plot in Figure~\ref{fig:sweeps_and_metrics}(a), the cases where the relative angle is positive -- meaning that there is a residual opening after the simulation finished running -- are coloured in green. We call this a state of under-closure. Cases for which the relative angle is approximately zero (more precisely: in the range $\left[-0.1,0.1\right]$), i.e. the cut is properly closed, are depicted in black. Finally, even negative values of $\alpha_{\text{final}} / \alpha_{\text{initial}}$ are possible in our simulations (coloured in purple). We call this state over-closure.
Over-closure can arise when the two cut edges of the simulation mesh move over or even through each other. This does not necessarily represent a biologically realistic scenario. 
For better visibility, we have separated the areas of under-, proper- and over-closure by white lines in the heatmap in Figure~\ref{fig:sweeps_and_metrics}(a). 
This representation can also be understood as a phase diagramme, separating the areas in the space of ($\lambda_{0}$, $\lambda_{r}$)-values that result in the different levels of closure. 

A visual representation of these levels is provided in Figure~\ref{fig:sweeps_and_metrics}(c). The three plots shown here are the final simulation states for the cases marked by the white symbols in Figure~\ref{fig:sweeps_and_metrics}(a) and using the same simulation mesh as in Figure~\ref{fig:linear_cut}. Here, the mesh points and springs at the edges of the cut lines are highlighted such that the level of closure is apparent. \\

We furthermore ask whether the phase diagramme depicted in Figure~\ref{fig:sweeps_and_metrics}(a) depends on the type of variation of the radial strain field, determined by the parameter $p$ in~\eqref{eqn:radial_strain_field}.

Figure~\ref{fig:sweeps_and_metrics}(d) therefore shows a set of averaged heatmaps for a range of different $p$-values (linear: $p=1$, quadratic: $p=2$, cubic: $p=3$, sqrt: $p=1/2$). 
As can be seen in these plots, in all cases the relative opening angle is smaller than $1$, indicating that the initiation of wound closure under radially contractile strain is independent of the shape of strain field variation. 
However, the individual heatmaps differ among one other.
That means for a given pair of strain values ($\lambda_{0}$, $\lambda_{R}$), the amount of closure achieved in the simulation is $p$-dependent. 
It follows that the precise location of the phase boundaries between under-, proper- and over-closed final states also depends on $p$. \\

We tested whether this phase boundary location might also depend on the choice of closure evaluation metric. So far, we have only considered the relative opening angle between final and initial states, $\alpha_{\text{final}} / \alpha_{\text{initial}}$. 
In the Supplementary Figure~\ref{fig:supp_sweeps_and_metrics} we therefore reproduced the results from Figure~\ref{fig:sweeps_and_metrics}(d), but now we have evaluated the amount of closure in different ways. In addition the the relative angle, we here used the relative 3D distance between the cut edge endpoints, $d_{\text{3D, final}} / d_{\text{3D, initial}}$, as well as the relative distance projected on the $xy$-plane, $d_{\text{2D, final}} / d_{\text{2D, initial}}$. 
As can be seen, the heatmaps in Figure~\ref{fig:supp_sweeps_and_metrics} are all very similar, indicating robustness among the evaluation metrics. We therefore continue to use the relative opening angle throughout the remainder of this work. \\

Finally, we also examine the averaged relative radii between the final and initial states for all the different strain and $p$-values considered here. As in Figure~\ref{fig:linear_cut}(e), Figure~\ref{fig:sweeps_and_metrics}(e) shows them as functions of $\lambda_{0}$ with different colours and symbols indicating different $\lambda_{R}$-values and locations of measurement. Interestingly, in the averages, the results for the umbrella rim and widest parts fall on top of each other even more strongly than for the one set of simulations shown in Figure~\ref{fig:linear_cut}(e).
Furthermore, the results for the different $p$-values differ in the strength of their dependence on $\lambda_{0}$ and $\lambda_{R}$. With increasing $p$, the $\lambda_{0}$-dependence becomes stronger (increasing slope of the linear fit), while the dependence on $\lambda_{R}$ appears the decrease. 
This indicates that the chosen higher contractility at the centre of the simulation mesh dominates the behaviour of the entire mesh in these cases.

\subsection*{The initiation of wound closure is also given for other cut shapes}

\begin{figure*}
    \centering
    \includegraphics[width=\textwidth]{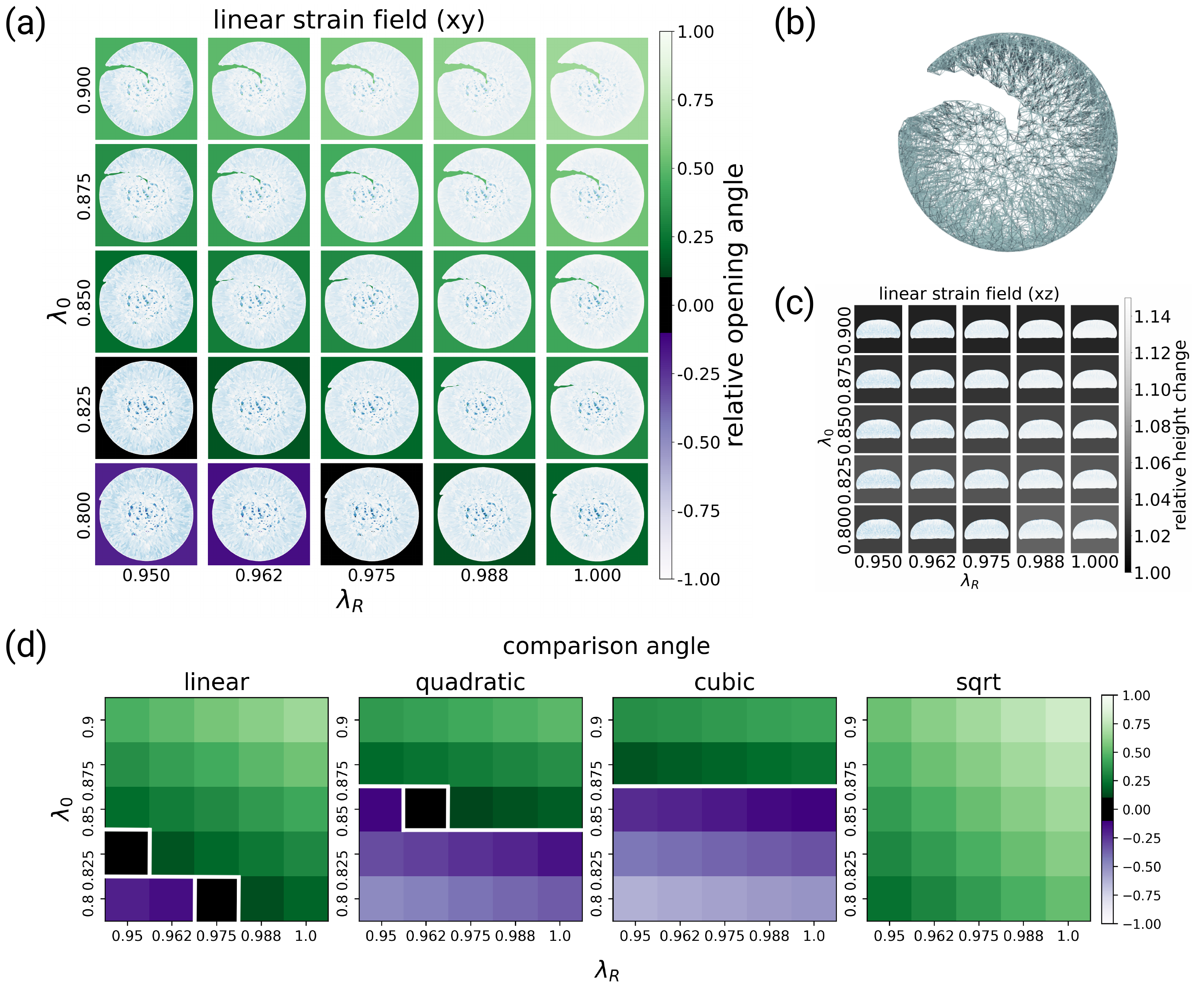}
    \caption{(a) Simulation results of wound closure initiation using a mesh with a $30\degree$ spiral cut under radial contractile strain fields (strain field parameters and colour bar same as in Figure~\ref{fig:linear_cut}(c)). 
    (b) Simulation mesh (initial state) with the $30\degree$ spiral cut in top view. 
    (c) Results of the same simulations as in (a). Here, the side views of the simulations' final states and relative height changes are shown (colour bar same as in Figure Figure~\ref{fig:linear_cut}(e)).
    (b) Comparison of amounts of closure for four parameter sweeps using the same $p$-values as in Figure~\ref{fig:supp_sweeps_and_metrics}(d). Note that here, only one simulation run has been performed for each pair of ($\lambda_{0}$, $\lambda_{R}$)-values. Hence, no averaging has taken place.}
    \label{fig:spiral_cut}
\end{figure*}

Next, it is crucial to ask whether wound closure can be initiated equally well for different wound shapes.

In particular, we compared our results obtained using the $\alpha_0 = 30\degree$ linear cut with results obtained using an $\alpha_0 = 30\degree$ logarithmic spiral cut, the simulation mesh's initial state (top view) of which is provided in Figure~\ref{fig:spiral_cut}(b). 

Figures~\ref{fig:spiral_cut}(a) and (c) show simulation results of runs with the same parameters as in Figure~\ref{fig:linear_cut}(c) and (e), but now obtained using the $30\degree$ spiral cut mesh. 
As the heatmaps show, all relative opening angles are smaller than $1$.
Again, this indicates that wound closure is initiated -- similarly to the linear cut case -- across the whole range of tested, radially contractile ($\lambda_{0}$, $\lambda_{R}$)-values. 

Furthermore, in Figure ~\ref{fig:spiral_cut}(d), we show the heatmaps for all the different strain field shapes (i.e. $p$-values) that we have considered. 
Although these heatmaps for the spiral cut shape are produced from individual runs (that is, they have not been averaged over multiple runs with different meshes) it can be seen, that they show similarities to the ones in Figure~\ref{fig:sweeps_and_metrics}(d).
This indicates that, while the amount of closure depends on the precise variation the of the strain field (i.e. on $p$), it might not strongly depend on the precise cut shape.

\subsection*{The amount of wound closure depends on the amount of remaining tissue material}

\begin{figure*}
    \centering
    \includegraphics[width=\textwidth]{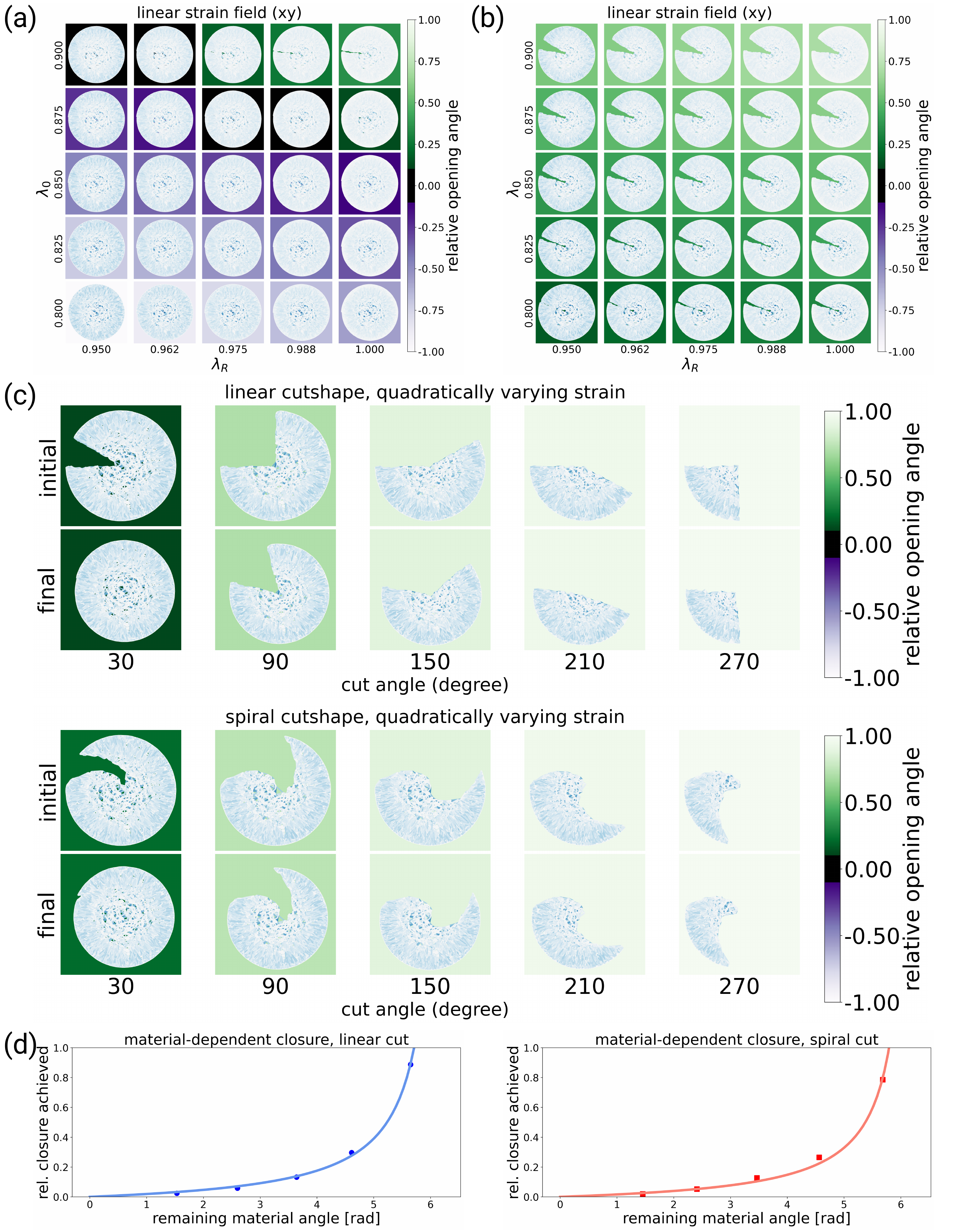}
    \caption{caption see next page}
    
\end{figure*}

\begin{figure*}[ht]
    \ContinuedFloat 
    \centering
    \caption{(a) Simulation results of wound closure initiation using a mesh with a $15\degree$ linear cut under radially contractile strain fields, with all other parameters and the colour bar same as in Figure~\ref{fig:linear_cut}(c). 
    (b) Simulation results of wound closure initiation using a mesh with a $45\degree$ linear cut under radially contractile strain fields, with all other parameters and the colour bar same as in Figure~\ref{fig:linear_cut}(c).
    (c) Examples of initial and final simulation stages for meshes with various cut angles and using a single, radially contractile strain field with $p=2$, $\lambda_{\text{centre}} = 0.875$ and $\lambda_{\text{edge}}=0.95$. Both linear cuts (top image) and spiral cuts (bottom image) have been included. Cut angles $\alpha_0$ are indicated at the $x$-axes of the plots. Colour bar same as in Figure~\ref{fig:linear_cut}(c).
    (d) Achievable relative closure as a function of remaining tissue material after cutting. Data points indicate the actual remaining material angles $\beta$ ($x$-axis, in radians) and actual achieved relative closures $1-\alpha_{\text{final}} / \alpha_{\text{initial}}$ ($y$-axis) for the simulations shown in (c) (left panel: linear cuts, right panel: spiral cuts). Here, $\beta$ is used as a proxy for the remaining tissue material. Lines represent the fits of~\eqref{eqn:closure_as_function_of_remaining_material} to the data. Obtained fit parameters: $c_{1} \approx 18.191$ and $c_{2} \approx 181.809$ for the linear cuts, $c_{1} \approx 15.561$ and $c_{2} \approx 184.438$ for the spiral cuts.}
    \label{fig:amounts_of_material}
\end{figure*}

Finally, we asked whether, and if so in which way, our previous results hold true for cuts with different cut angles.

Figures~\ref{fig:amounts_of_material}(a) and (b) show two sets of simulation runs for $\alpha_0 = 15\degree$ and $\alpha_0 = 45\degree$ linear cuts. 
Comparing these two plots and the one in Figure~\ref{fig:linear_cut}(c), shows that, for the same strain field, the amount of closure achieved in the simulation strongly depends on the initial size of the cut. For example, while the strain field with $\lambda_{0}=0.825$ and $\lambda_{R}=0.975$ results in almost perfect closure for the mesh with $\alpha \approx 30\degree$, the same strain field lead to over-closure for $\alpha \approx 15\degree$, and under-closure for $\alpha \approx 45\degree$.
Thus, the amount of achievable closure not only depends on the strain field itself, but also on the initial size of the wound.  

We therefore wondered what would happen if the cut was very large and hence the amount of remaining tissue material very small. 
Figure~\ref{fig:amounts_of_material}(c) shows the results of simulations with one specific strain field (namely $p=2$, $\lambda_{0} = 0.875$ and $\lambda_{R}=0.95$) and varying cut angles for both linear (top image) and spiral (bottom image) cuts. The values of the initial cut angles are provided on the $x$-axes of the plots.
As can be seen from this figure, the relative amount of closure determined $\alpha_{\text{final}} / \alpha_{\text{initial}}$ indeed strongly decreases with increasing initial cut angle.
However, even for large cuts (e.g. $\alpha \approx 210\degree$) there is a tendency of the system to reduce the cut opening in the simulation. 
Thus, even small residual pieces of the mesh are capable of initiating wound closure in our simulation. \\

Note that in the left-most panel of the plots for the linear cut shapes in Figure~\ref{fig:amounts_of_material}(c), the background colour indicates that the cut is under-closed while the mesh (in top view) appears to be proper-closed. The reason for this seeming discrepancy is that, while a cut can be proper-closed across nearly the entire cut line, it might remain slightly open very close to the rim. For a visual representation of this effect, we have plotted the same simulation's final state in a different way in Supplementary Figure~\ref{fig:supp_amounts_of_material}(a). In this case, only cross-layer connections and the cut line are shown in side-view. Here, the residual opening of the cut very close to the umbrella rim is apparent. Due to the way we calculate the opening angles $\alpha$ from the cut edge points close to the rim, this effect influences the final opening angle $\alpha_{\text{final}}$ and hence the classification of a given simulation into under-, proper-, or over-closed levels. However, we found that the influence of this effect is generally small (compare Supplementary Figure~\ref{fig:supp_amounts_of_material}(b)) and therefore do not further account for it. \\

It is indeed even possible to estimate the amount of closure as a function of the remaining amount of material analytically. 
To this end, we concentrated on the $xy$-plane projection of umbrella rim and considerd its shape changes throughout a strain-induced deformation.

Before the strain field is applied, the rim has radius $\mathcal{R}$ and circumference $U = 2 \pi \mathcal{R}$. With a given cut angle $\alpha$, the arc length of the cut-away piece is $A = \alpha \mathcal{R}$ and thus the arc length of the remaining piece of tissue is $B = U - A = \beta \mathcal{R}$, with a remaining angle $\beta = 2 \pi - \alpha$. 

During the deformation, the tissue shrinks in the radial direction and expands circumferentially (however, the following arguments are equally valid for radially expanding and circumferentially shrinking tissues).

After the shape change, the umbrella rim has adopted a new radius $\rho$ (in our case $\rho \leq \mathcal{R}$). A full circle with this radius would thus have a circumference of $u = 2 \pi \rho$. Simultaneously, the tissue material on the rim has also deformed circumferentially, changing length from $B$ to $kB$. Importantly, this new material length constitutes the rim tissue in the deformed configuration. Thus, the residual cut opening now has an arc length of $a = u-kB$ and is associated to a new, residual opening angle
\begin{equation}
    \alpha' = \frac{a}{\rho} = \frac{u-kB}{\rho} = \frac{2 \pi \rho - k \beta \mathcal{R}}{\rho} \, .
\end{equation}
This means, the relative opening angle between the final and initial stages i.e. the quantity we have been plotting throughout this work reads
\begin{equation}
    \frac{\alpha_{\text{final}}}{\alpha_{\text{initial}}} = \frac{\alpha'}{\alpha} = \frac{2 \pi \rho - k \mathcal{R} \beta}{\alpha \rho} \, .
\end{equation}
Expressing the right-hand side of this last expression solely in terms of $\beta$, which we here used as a proxy for the remaining amount of material ($\beta$ indeed accurately represents the remaining amount of material at least in the linear cut case),
\begin{equation}
    \frac{\alpha'}{\alpha} = \frac{2 \pi \rho - k \mathcal{R} \beta}{(2 \pi - \beta) \rho} \, .
\end{equation}
Finally, since we aimed to understand the relative achievable closure dependent on the residual amount of material, we also calculated $1 - \alpha'/\alpha$, which is zero if the opening angle does not change (i.e. $\alpha' = \alpha$) and one if it is fully closed after the strain-induced deformation (i.e. $\alpha' = 0$). This yields
\begin{equation}
    1 - \frac{\alpha'}{\alpha} = 1 - \frac{2 \pi \rho - k \mathcal{R} \beta}{(2 \pi - \beta) \rho} = \beta \frac{k \mathcal{R} - \rho}{2 \pi \rho - \rho \beta} \, .
    \label{eqn:closure_as_function_of_remaining_material}
\end{equation}
The last equality contains two unknown variables, $c_{1} = k \mathcal{R} - \rho$ and $c_{2} = \rho$, which can be obtained from fitting the data in Figure~\ref{fig:amounts_of_material}(d) (for details of the fitting procedure, see Methods ``Fitting the amount of closure as a function of remaining tissue material'').
In both plots in Figure~\ref{fig:amounts_of_material}(d), the data from the simulations Figure~\ref{fig:amounts_of_material}(c) is shown together with the fit of~\eqref{eqn:closure_as_function_of_remaining_material}.
The fitting parameters are $c_{1} \approx 18.191$ and $c_{2} \approx 181.809$ for the linear cut, and $c_{1} \approx 15.561$ and $c_{2} \approx 184.438$ for the spiral cut shape. 
As can be seen, these fits successfully explain the data. 

\section*{Discussion} \label{sec:Discussion}

In summary, in this paper, we took inspiration from experimental results demonstrating rapid wound closure in the jellyfish \textit{Clytia hemisphaerica}~\cite{Sinigaglia2020} and from the possibility that this process might be driven mechanically. 

We first introduced an idealised \textit{in silico} model of the \textit{Clytia} umbrella.
We then employed this model together with a broad range of strain fields to examine the possibility that pre-strains in the umbrella act as initiators and drivers of wound closure in jellyfish systems. 

Our simulations showed that radially contractile but not radially extensile strain fields indeed lead to an initiation of wound closure in our model. 
This observation forms the basis of our entire subsequent analysis and is therefore fundamental both in the context of this work and future, more detailed studies of jellyfish wound closure. 
The result that radially contractile strains lead to cut closure might not seem unexpected, as common elastic materials exhibit the Poisson effect, meaning that, under strain, they not only deform in the direction of loading but also perpendicular to it. 
For radially contractile strains this also implies a circumferential deformation, leading to changes in cut opening angles. Such effects have been studied in simpler material geometries before, for example in~\cite{Modes2013}. 
However, given that even initially flat shape programmable materials are known to display very surprising and unexpected deformations under application of apparently simple strain fields~\cite{Modes2012, Yau2026}, it is of great importance to verify that results obtained in simpler geometries still hold true in more complicated ones. 
Thus, our results presented here extend previous work to the more complex, naturally occurring shape of a jellyfish body and provides important insights for the field of shape programmable materials. 
In this regard, it will also be interesting to investigate in more detail why we observe an increase in height of the simulation mesh under both radially contractile and radially extensile strain fields, even though these two types of strain lead to opposite behaviours in terms of radius change and wound closure (or opening). 

After having established that it is indeed possible to initiate wound closure via simple, radially contractile pre-stains in our system, we next quantified the amount of achievable closure in more detail. 
Based on our chosen strain field expression, we considered the dependence on strain field magnitudes $\lambda_{0}$ at the umbrella apex, and $\lambda_{R}$ at the widest part of the umbrella, as well as on the variation between the two, i.e. the parameter $p$. 
We found three levels of closure -- which we named under-, proper- and over-closure -- and discovered that the phase diagramme of these levels in the space of ($\lambda_{0}$, $\lambda_{R}$)-values strongly depends on $p$ and on the wound opening angle $\alpha$. 
Remarkably, the phase diagramme appears to be less sensitive to the precise shape of the cut lines (straight or spiral), when $p$ and $\alpha$ are kept constant. 
Because we here only studied simple types of cuts, it will be important to consider a wider range of cut shapes in the future.  
Experimentally, wound closure has been observed across a huge variety of different cuts~\cite{Sinigaglia2020} and it is compelling to re-create these observations \textit{in silico}.

Nevertheless, with our simple selection of cut shapes, we have shown that, while the amount of achievable closure strongly depends on the initial size of the cut, even small pieces are still able to initiate closure. 
This result is in good accordance with previous experiments~\cite{Sinigaglia2020}.  
Beyond that, we were able to capture the functional form of the dependence between amount of remaining material (approximated by the remaining material angle $\beta$) and the closure achievable in our simulation. 
This underlines that even very simple models can capture fundamental properties of a complex system. 

However, in order to achieve full agreement between the \textit{in silico} model and the biological system, further adjustments are of course going to be necessary. 

From the modelling side, so far, we did not consider the effects of our chosen strain fields on the healthy, uncut jellyfish umbrella. Ideally, in the uncut state the strain would be balanced in such a way that it keeps the umbrella in its `natural' shape and only initiates radial contraction and thus wound closure once the cut is being administered. 
However, for this balance to be in place, in reality multiple tissue layers interact. In particular, this includes the contractile epitheliomuscular cells~\cite{Leclere2017}, i.e. the muscles of the jellyfish, and the thick mesoglea. 
Through their architecture and arrangement, both of these structures likely contribute in non-trivial ways to overall umbrella shape maintenance and swimming movements. 
In terms of muscle, both radial and circumferential fibres exist, of which, additionally, the former are smooth and the latter striated. Thus, it can be expected that complex contraction patterns are possible, as are indeed seen during changes in swimming direction~\cite{Gladfelter1973}. 
In terms of the mesoglea, we are currently examining its properties in more detail. But even the most apparent feature -- its changing thickness between umbrella apex to periphery -- is likely to be an important player in the finely balanced interplay of shape and movement. 
Hence, the strain fields we have been applying in our simulations can only be understood as aggregate patterns emerging across the entirety of the umbrella structure. 
If we aim for more realistic strains, than different tissue layers necessarily have to be considered explicitly.

Further anatomical features of the jellyfish body have also not been included in our \textit{in silico} model so far. For example, around the gonads or the manubrium, the tissue layers might have different elastic properties than elsewhere, resulting in changes in the effects of homogeneous strains or even in spatial variations of the strain fields themselves. 

All of these aspects show that there is still some way to go from our minimal model to a full \textit{in silico} representation of a \textit{Clytia} jellyfish.
Yet, even in its minimal state, our model provides important new insights into the underlying mechanisms of jellyfish wound closure.
Most importantly, in this work, we showed that it is indeed possible that mechanical processes initiate wound closure in \textit{Clytia hemisphaerica}.
Furthermore, we investigated the influence of individual factors such as strain field parameters and wound size on the amount of achievable closure. 
Finally, we provided an analytical expression for the amount of mechanically achievable closure depending on the remaining amount of tissue material.

It is now intriguing to link all of these results back to experiments and confirm our findings in the real biological setting. 
In particular, it will be interesting to measure the actual underlying strain fields in the jellyfish umbrella and confirm whether they are indeed radially contractile as predicted in this work. 
Furthermore, it will be fascinating to see whether our prediction regarding the amount of closure as a function of residual tissue material really holds true. To test this, the dynamics of wound closure after administering cuts of different sizes might need to be followed experimentally. 
Finally, also considering both the healthy (uncut) organism and processes at the cell scales will provide clues about the relative contributions of cellular vs. mesogleal layers. 
The combination of experimental and modelling work thus constitutes a powerful tool to obtain an in depth insight into the fundamental processes of wound healing, but also propulsion, in the Cnidarian branch of the animal tree of life.

\section*{Acknowledgments}
The authors would like to thank Daniel Font-Martín and Ulyana Shimanovich for helpful discussions on this work. A.M. and Z.S. would also like to thank Lior Moneta for sharing pieces of code that were adapted in the surface sampling and mesh generation procedures. Furthermore, A.M. and Z.S. would like to thank Wan Yee Yau and Lior Moneta for helpful discussions concerning the spring lattice simulations.
A.M. acknowledges support by the Human Frontier Science Program (HFSP) under research grant 2024 1154043. Z.S. acknowledges support through the
International Max Planck Research School for Cell, Developmental and Systems Biology (IMPRS-
CellDevoSys) Student Summer Research Internship program.

\section*{References}
\bibliographystyle{unsrt}
\bibliography{references.bib}

\section*{Methods} \label{sec:Methods}

\subsection*{Geometric description of an idealised jellyfish body} 

For our simulations, we first built a 3-dimensional representation of a cut jellyfish umbrella (compare Figure~\ref{fig:intro_problem_setup}(d) left part). Our representation is idealised because we do not distinguish any tissue layers and consider the (uncut) umbrella to be perfectly rotationally symmetric, omitting the existence of canals and gonads, for example.

We outlined the cross-sectional shapes of the exumbrella (outer) and subumbrella (inner) surfaces of a jellyfish body in a video capture image of a swimming (uninjured) \textit{Clytia hemisphaerica} (compare also Figure~\ref{fig:intro_problem_setup}(a)) by manually selecting points on the jellyfish tissue surfaces under high image magnification (magnification: $327 \%$, image rotated by $122.5\degree$ to align body symmetry axis in $z$-direction). 
We then fitted these sample points using the \textit{polynomial.polynomial.Polynomial.fit()} function of the NumPy package in Python~\cite{Harris2020} to obtain an $r(z)$-representation, with $r$ the radial distance of the surface from the jellyfish centre line and $z$ the height of a point above the jellyfish rim. Fitting $r(z)$, not $z(r)$, was necessary because some jellyfish umbrellas are actually wider away from the rim than at the rim itself, hence fitting $z(r)$ would not have resulted in a proper functional expression. 
The polynomial fit was chosen because of its simplicity and tractability. Testing multiple polynomial degrees and outlines of other species (\textit{Pandea rubra}~\cite{PandeaImage}, \textit{Aurelia aurita}~\cite{AureliaImage}) obtained from open source web images, revealed that the smallest degree that faithfully represents general jellyfish shapes is $6$.
However, since the \textit{Clytia hemisphaerica} umbrella is very flat, the $6^{\text{th}}$-degree polynomial was joined to a small linear piece capturing the apex. 
We thus obtained the following functions for the \textit{Clytia hemisphaerica} mesoglea shape (in arbitrary units):
\begin{strip}
\begin{equation}
    r_{out}(z)=
    \begin{cases}
       \parbox{0.75\textwidth}{
       $- 29.34077495 (-1.0 + 0.01263424 z)^6$  
        $- 27.62218143 (-1.0 + 0.01263424 z)^5$ \\
        $+ 4.06697313 (-1.0 + 0.01263424 z)^4$ 
        $+ 4.09720864 (-1.0 + 0.01263424 z)^3$ \\
        $- 40.6028472 (-1.0 + 0.01263424 z)^2$ 
        $- 40.21779324 (-1.0 + 0.01263424 z)$ 
        $+197.94420312$} & z < 158.3 \\
      -\frac{68.3247}{3.4}z + \frac{68.3247}{3.4}  \cdot 161.7 & z \geq 158.3
   \end{cases}
\end{equation}
\begin{equation}
    r_{in}(z)=
    \begin{cases}
       \parbox{0.75\textwidth}{
       $- 29.83670765 (-1.0 + 0.01594896 z)^6$
       $- 30.98626065 (-1.0 + 0.01594896 z)^5$ \\
       $+ 16.91461828 (-1.0 + 0.01594896 z)^4$
       $+ 27.45021172 (-1.0 + 0.01594896 z)^3$ \\
       $- 40.98800364 (-1.0 + 0.01594896 z)^2$
       $- 58.90941428 (-1.0 + 0.01594896 z)$
       $+ 179.15033476$
       } & z < 125.4 \\
      -\frac{62.7949}{2.4}z + \frac{62.7949}{2.4}  \cdot 127.8 & z \geq 125.4
   \end{cases}
\end{equation}
\end{strip}
For the purpose of this work, our jellyfish representation then was the body of revolution around the $z$-axis of the area enclosed by these curves.

\subsection*{Point sampling and mesh generation}
Next, we randomly sampled multiple sets of points on the outer and inner surfaces of this body of revolution, using the extension of a sampling procedure previously described in~\cite{Modes2008} (compare Figure~\ref{fig:intro_problem_setup}(d) middle part). 
We further obtained mathematical expressions for straight and spiral cut lines and examined which of the sampled points lay in the cut-away region in each case. 
Then, we triangulated each individual surface, using the \textit{geometry.TriangleMesh.create\_from\_point\_cloud\_ball\_pivoting()} function from the Open3D Python package~\cite{Zhou2018}. Points in the cut-away regions were being removed from the meshes only after triangulation to avoid artificial straightening of the meshes' cut lines. 
Finally, surfaces were being linked by cross-connecting individual points on each side with the three nearest points of the other side~\cite{Ramos2025}.
With this, we obtained our simulation meshes as uniformly sampled \textit{in silico} representations of the idealised jellyfish umbrella, as also shown in Figure~\ref{fig:linear_cut}(a).

\subsection*{Material properties of the \textit{in silco} jellyfish mesoglea}
As we performed a purely theoretical study of the general conditions leading to wound opening or closure in the jellyfish body geometry, we did not make any assumptions about the underlying material properties of the tissue. Therefore, in the simulations, we set all the material parameters, including spring constants and the mass distribution on the mesh points, to $1$. 

\subsection*{Strain fields}
We here describe how the strain fields defined in~\eqref{eqn:strain_tensor_field} and~\eqref{eqn:radial_strain_field} have being applied to the simulation meshes. 
In order to do so, following~\cite{Fuhrmann2024}, we first discretised $\boldsymbol{\lambda} (\boldsymbol{x})$ for each spring. 
For a spring connecting two vertices $j$ and $k$ with positions $\boldsymbol{x}_{j}$ and $\boldsymbol{x}_{k}$ we evaluated the strain tensor at the two endpoints and took the average value as the approximation,
\begin{equation}
\boldsymbol{\lambda}_{jk}=\frac{1}{2}\left[\boldsymbol{\lambda} (\boldsymbol{x}_{j}) +\boldsymbol{\lambda} (\boldsymbol{x_{k}})\right].
\label{eqn:strain_field_for_spring}
\end{equation}
For a spring with initial length
\begin{equation}
\delta_{\text{initial}}=\left\| \boldsymbol{x}_{j} -  \boldsymbol{x}_{k}\right\|=\left\| \Delta \boldsymbol{x} \right\| \, ,
\label{eqn:spring_ini_length}
\end{equation}
the target length thus takes the form
\begin{equation}
\delta_{\text{target}}=\left\|\boldsymbol{\lambda}_{ij}  \Delta \boldsymbol{x} \right\| \, ,
\label{eqn:spring_fin_length}
\end{equation}
(compare Figure~\ref{fig:intro_problem_setup}(c)). 
During a simulation, the springs relax towards their target lengths through overdamped dynamics, following the details in~\cite{Fuhrmann2024}.
However, individual springs might not reach their target lengths in the simulation runs, leaving residual strains in the system.  

\subsection*{Simulation runs}
To run our simulations, we used the spring lattice model implementation of our previously published software package TopoSPAM~\cite{Singh2025, TopoSPAM} (for other use cases of this spring lattice implementation, see for example~\cite{Fuhrmann2024, Ramos2025, Yau2026}). 
In the simulation runs, we set the time stepping parameter to $dt=0.01$ and the tolerance parameter to $tol=10^{-6}$. The latter is compared to the average movement of all the points during a simulation time step. The simulation thus terminates once the average point movement is smaller than $tol$. Hence, it is possible that residual strains remain in the system.

\subsection*{Quantification of wound closure}
To evaluate the amount of closure achieved by a given simulation, we analyzed the relative changes of the cut angle $\alpha$ and cut opening distance $d$. 
To this end, we defined three reference points on each surface of our simulation mesh: the cut centre $\boldsymbol{x}_{\text{centre}}$ where the two cuts lines meet close to the umbrella apex, and the left and right cut endpoints, $\boldsymbol{x}_{\text{left}}$ and $\boldsymbol{x}_{\text{right}}$, at the umbrella rim. These reference points were extracted from the mesh points that lie on the edge of the cuts in the initial configuration (i.e. before the simulation is run).
We then traced the IDs of the reference points throughout the simulation to obtain their positions in the final state.

In more detail, to obtain the cut endpoints we considered all the points on the cut edges that are close to the umbrella rim. That is, we selected all those with $z$-coordinates lying in the interval $\left[0.02z_{\text{max}}, 0.2z_{\text{max}}\right]$, with $z_{\text{max}}$ being the maximum value of mesh points' $z$-coordinates. We also re-ran our entire analysis with a different interval, namely $\left[0.02z_{\text{max}}, 0.5z_{\text{max}}\right]$, for comparison (see Supplementary Figure~\ref{fig:supp_amounts_of_material}(b)). 
We then classified the selected points with respect to the angular bisector of the cut, labelling points as either 'left' or 'right'. 
Then, we defined $\boldsymbol{x}_{\text{left}}$ and $\boldsymbol{x}_{\text{right}}$ to be the averaged positions of the two labelled sets of points. 
For $\boldsymbol{x}_{\mathrm{centre}}$, since the mesh is discrete in nature and might not contain a point with $z$ precisely $z=0$, we chose the mesh point on the cut edges that is closest to the $z$-axis.

With these definitions, the cut angle $\alpha$ is the angle between the vectors $\boldsymbol{x}_{\text{left}}-\boldsymbol{x}_{\text{centre}}$ and $\boldsymbol{x}_{\text{right}}-\boldsymbol{x}_{\text{centre}}$ projected onto the $xy$-plane. We called these projected vectors $\boldsymbol{v}_{\text{L}}$ and $\boldsymbol{v}_{\text{R}}$. 
Using them, we first computed an angle $\phi$ as
\begin{equation}
\phi
=\arccos \left( \frac{\boldsymbol{v}_{\text{L}}\cdot \boldsymbol{v}_{\text{R}}}{|\boldsymbol{v}_{\text{L}}|| \boldsymbol{v}_{\text{R}}|} \right) \, , \quad\phi \in[0,\pi].
\label{vector_angle}
\end{equation}
Since this angle alone is not sufficient to distinguish between pieces of the mesh smaller and larger than half, we combined it with the sign of 
\begin{equation}
C=\mathbf{e}_z\cdot
\left(
\boldsymbol{v}_{\text{L}}\times \boldsymbol{v}_{\text{R}}
\right) \, ,
\label{vector_cross_porduct}
\end{equation}
where $\mathbf{e}_z$ is the unit vector along the z-axis, and with the theoretical angle $\alpha_0$ that we originally used to define the cut lines for generating the mesh. Then, we obtained the real cut angle $\alpha$ on the mesh as 
\begin{equation}
\alpha
=
\begin{cases}
\phi, 
& C<0,\\[4pt]
2\pi-\phi,
& C\geq 0 \ \text{and}\ \alpha_0\geq \pi,\\[4pt]
-\phi,
& C\geq 0 \ \text{and}\ \alpha_0<\pi.
\end{cases}
\label{cut_angle}
\end{equation}
In this way, the cut angle of an actual open cut is defined as $\alpha\in[0,2\pi)$. Negative $\alpha$-values represent the over-closed state, in which parts of the mesh have moved on top of or through each other.

Furthermore, for comparison, we also quantified the 3D opening distance $d_{\text{3D}}$, i.e. the distance between $\boldsymbol{x}_{\text{left}}$ and $\boldsymbol{x}_{\text{right}}$, and 2D opening distance $d_{\text{2D}}$, being the distance projected onto the $xy$-plane.  
Again, we used the signed values of $d$ to distinguish the final configurations with residual openings and overlapping endpoints. 

Unless stated otherwise, we report the relative change of the opening angle (or distance), i.e. the value $\alpha_{\text{final}}/\alpha_{\text{initial}}$ (or $d_{\text{final}} / d_{\text{initial}}$). 
Since the thickness change of the mesh is negligible, we averaged the values of inner and outer mesh surfaces.

\subsection*{Quantification of mesh shape change}
Furthermore, to investigate how the strain fields deform the whole mesh shape, we quantified its relative height and radius changes. 
The height is defined as the difference $h=z_{\text{max}}-z_{\text{min}}$ and computed separately for the two surfaces, and then averaged. 
For the radius change, we considered two separate regions: one is the rim of the mesh, and the other its widest part. 
We defined these regions as containing all the points with $z\in[z_{\min},\,z_{\min}+2\Delta z]$ and with $z\in[z^*-\Delta z,\,z^*+\Delta z]$ where $z^*$ is the $z$-coordinate of the point with maximum radius, respectively. The region width was set to 2$\Delta z$, with $\Delta z=0.03z_{\text{max}}$ in both cases. 
For each region, we calculated its average radius and then averaged again over the two mesh surfaces. 
As for the opening angle (and distances), we traced the reference points' IDs from the initial to the final state of the simulation and report the relative changes of mesh height and radii with respect to their initial values.

\subsection*{Averaging over multiple simulation runs}
In some cases, we pooled multiple simulations with the same strain field but different simulation meshes, to obtain average closure values and standard derivations thereof or average radii changes. 
In these cases, we calculated all the quantities separately for each individual simulation run and averaged over runs afterwards.

\subsection*{Fitting the amount of closure as a function of remaining tissue material}
To fit the data of the amount of closure as a function of the remaining tissue material to~\eqref{eqn:closure_as_function_of_remaining_material} derived in the main text, we used the \textit{optimize.curve\_fit()} function of the SciPy package in Python~\cite{Vugrin2007, Virtanen2020} with $c_{1}$ and $c_{2}$ being the two variables to be fitted. Additionally, fit bounds $0 \leq c_{1} \leq 100$ and $100 \leq c_{2} \leq 200$ were provided, because the structure of~\eqref{eqn:closure_as_function_of_remaining_material} allows for scalings between these two unknowns. The bounds were chosen based on the facts that the maximum mesh radius has a value of approx. $200$ in arbitrary units before the strain field is being applied and that it shrinks to about $90\%$ of its initial value during the simulations (compare Figure~\ref{fig:sweeps_and_metrics}(e)).

\section*{Supplementary Figures}

see next page

\begin{figure*}
    \centering
    \includegraphics[width=\textwidth]{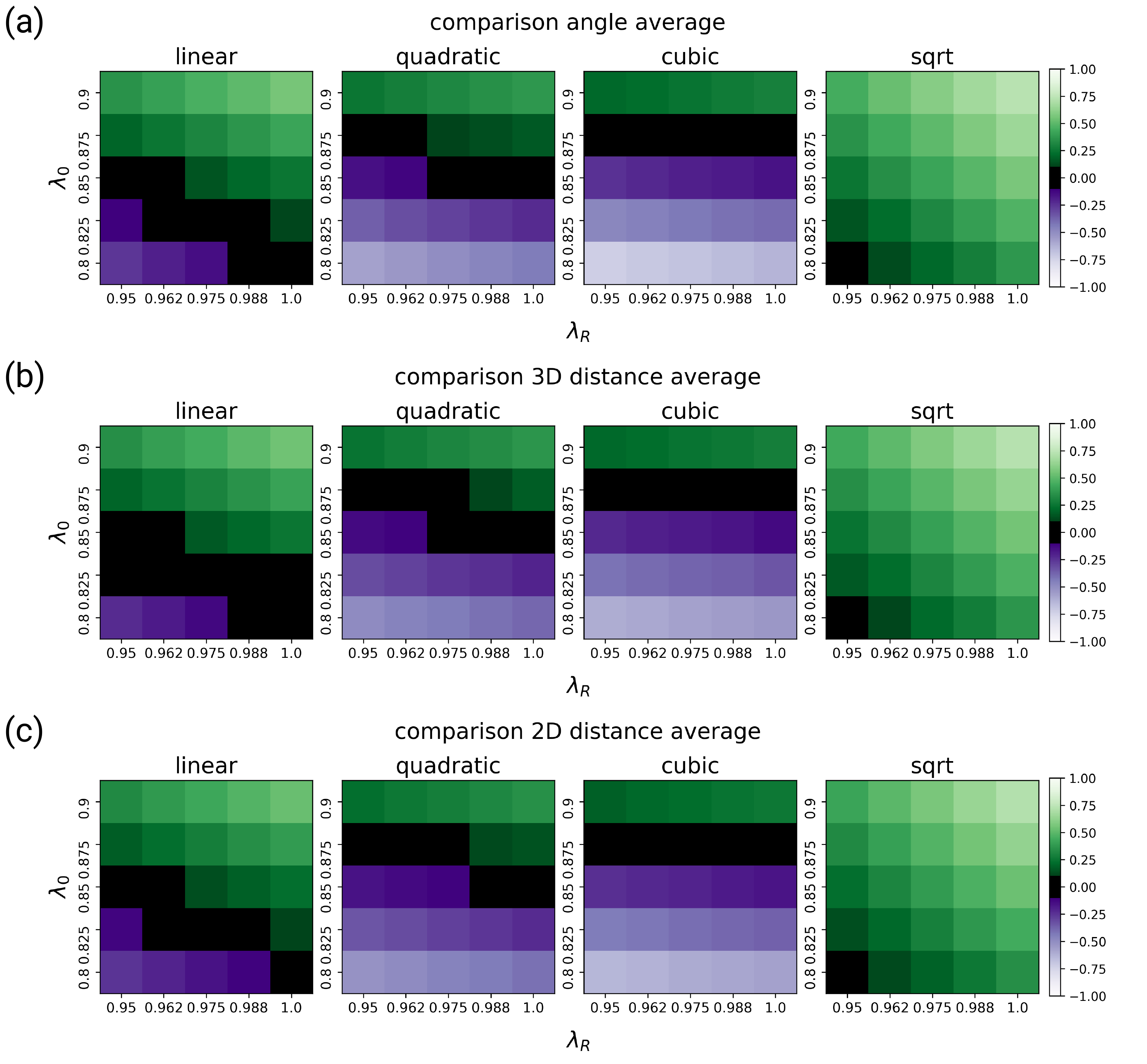}
    \caption{Comparison of evaluation metrics. (a) Same as in Figure~\ref{fig:sweeps_and_metrics}(d). 
    (b) Same simulation results as in (a), but instead of the averaged relative opening angles between final and initial stages, these plots show the averaged relative 3D opening distances $d_{\text{3D, final}} / d_{\text{3D, initial}}$.
    (c) Same simulation results as in (a), but instead of the averaged relative opening angles, these plots show the averaged relative 2D opening distances $d_{\text{2D, final}} / d_{\text{2D, initial}}$.}
    \label{fig:supp_sweeps_and_metrics}
\end{figure*}

\begin{figure*}
    \centering
    \includegraphics[width=\textwidth]{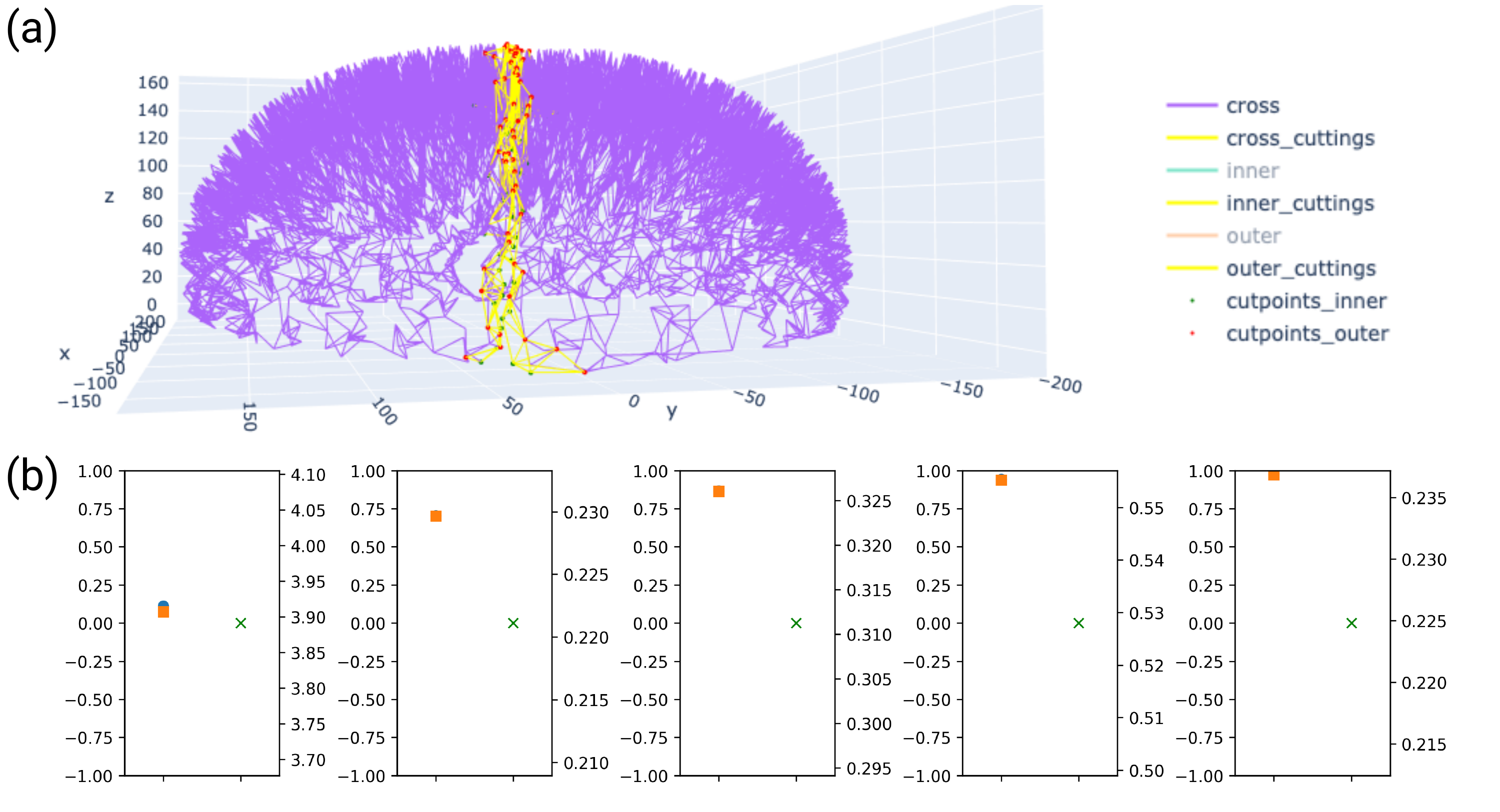}
    \caption{(a) Same simulation result as in Figure~\ref{fig:amounts_of_material}(c), left-most panel of simulations with linear cuts, in a different view. Here, only the mesh's cross connections (purple) and the cut edges (yellow) are shown in side-view, making the residual opening of the cut very close to the umbrella rim apparent. 
    (b) Estimation of the influence of such residual openings close to the umbrella rim on the calculation of the overall amount of closure. Here, we have repeated the calculation of the opening angles $\alpha_{\text{initial}}$ and $\alpha_{\text{final}}$ for the simulations with linear cuts shown in Figure~\ref{fig:amounts_of_material}(c). However, we included more cut edge points in the averaging for the tips, changing the intervals from $\left[0.02z_{\text{max}}, 0.2z_{\text{max}}\right]$ to $\left[0.02z_{\text{max}}, 0.5z_{\text{max}}\right]$. Blue and orange data points represent the quantity $\alpha_{\text{final}} / \alpha_{\text{initial}}$ with the old and new tip definitions, respectively (left hand side $y$-axes of the plots). Green crosses indicate the relative change of the resulting value (i.e. $(\alpha_{\text{final, old}} - \alpha_{\text{final, new}}) / \alpha_{\text{initial}}$) in percent (right hand side $y$-axes of the plots).}
    \label{fig:supp_amounts_of_material}
\end{figure*}

\end{document}